
%
%
%
\def\unredoffs{} \def\redoffs{\voffset=-.31truein\hoffset=-.59truein}
\def\speclscape{\special{ps: landscape}}
%
%
%
%
\newbox\leftpage \newdimen\fullhsize \newdimen\hstitle \newdimen\hsbody
\tolerance=1000\hfuzz=2pt
\catcode`\@=11 
\def\bigans{b }
\message{ big or little (b/l)? }\read-1 to\answ
\ifx\answ\bigans\message{(This will come out unreduced.}
\magnification=1200\unredoffs\baselineskip=16pt plus 2pt minus 1pt
\hsbody=\hsize \hstitle=\hsize 
\else\message{(This will be reduced.} \let\l@r=L
\magnification=1000\baselineskip=16pt plus 2pt minus 1pt \vsize=7truein
\redoffs \hstitle=8truein\hsbody=4.75truein\fullhsize=10truein\hsize=\hsbody
\output={\ifnum\pageno=0 
  \shipout\vbox{\speclscape{\hsize\fullhsize\makeheadline}
    \hbox to \fullhsize{\hfill\pagebody\hfill}}\advancepageno
  \else
  \almostshipout{\leftline{\vbox{\pagebody\makefootline}}}\advancepageno
  \fi}
\def\almostshipout#1{\if L\l@r \count1=1 \message{[\the\count0.\the\count1]}
      \global\setbox\leftpage=#1 \global\let\l@r=R
 \else \count1=2
  \shipout\vbox{\speclscape{\hsize\fullhsize\makeheadline}
      \hbox to\fullhsize{\box\leftpage\hfil#1}}  \global\let\l@r=L\fi}
\fi
%
\newcount\yearltd\yearltd=\year\advance\yearltd by -1900

\def\Title#1#2{\nopagenumbers\abstractfont\hsize=\hstitle\rightline{#1}%
\vskip 1in\centerline{\titlefont #2}\abstractfont\vskip .5in\pageno=0}
\def\Date#1{\vfill\leftline{#1}\tenpoint\supereject\global\hsize=\hsbody%
\footline={\hss\tenrm\folio\hss}}
%

\def\draftmode{\message{ DRAFTMODE }\def\draftdate{{\rm preliminary draft:
\number\month/\number\day/\number\yearltd\ \ \hourmin}}%
\headline={\hfil\draftdate}\writelabels\baselineskip=20pt plus 2pt minus 2pt
 {\count255=\time\divide\count255 by 60 \xdef\hourmin{\number\count255}
  \multiply\count255 by-60\advance\count255 by\time
  \xdef\hourmin{\hourmin:\ifnum\count255<10 0\fi\the\count255}}}
\def\nolabels{\def\wrlabeL##1{}\def\eqlabeL##1{}\def\reflabeL##1{}}
\def\writelabels{\def\wrlabeL##1{\leavevmode\vadjust{\rlap{\smash%
{\line{{\escapechar=` \hfill\rlap{\sevenrm\hskip.03in\string##1}}}}}}}%
\def\eqlabeL##1{{\escapechar-1\rlap{\sevenrm\hskip.05in\string##1}}}%
\def\reflabeL##1{\noexpand\llap{\noexpand\sevenrm\string\string\string##1}}}
\nolabels
%
\global\newcount\secno \global\secno=0
\global\newcount\meqno \global\meqno=1
\def\newsec#1{\global\advance\secno by1\message{(\the\secno. #1)}
\global\subsecno=0\eqnres@t\noindent{\bf\the\secno. #1}
\writetoca{{\secsym} {#1}}\par\nobreak\medskip\nobreak}
\def\eqnres@t{\xdef\secsym{\the\secno.}\global\meqno=1\bigbreak\bigskip}
\def\sequentialequations{\def\eqnres@t{\bigbreak}}\xdef\secsym{}
\global\newcount\subsecno \global\subsecno=0
\def\subsec#1{\global\advance\subsecno by1\message{(\secsym\the\subsecno.
#1)}
\ifnum\lastpenalty>9000\else\bigbreak\fi
\noindent{\it\secsym\the\subsecno. #1}\writetoca{\string\quad
{\secsym\the\subsecno.} {#1}}\par\nobreak\medskip\nobreak}
\def\appendix#1#2{\global\meqno=1\global\subsecno=0\xdef\secsym{\hbox{#1.}}
\bigbreak\bigskip\noindent{\bf Appendix #1. #2}\message{(#1. #2)}
\writetoca{Appendix {#1.} {#2}}\par\nobreak\medskip\nobreak}
%
%
\def\eqnn#1{\xdef #1{(\secsym\the\meqno)}\writedef{#1\leftbracket#1}%
\global\advance\meqno by1\wrlabeL#1}
\def\eqna#1{\xdef #1##1{\hbox{$(\secsym\the\meqno##1)$}}
\writedef{#1\numbersign1\leftbracket#1{\numbersign1}}%
\global\advance\meqno by1\wrlabeL{#1$\{\}$}}
\def\eqn#1#2{\xdef #1{(\secsym\the\meqno)}\writedef{#1\leftbracket#1}%
\global\advance\meqno by1$$#2\eqno#1\eqlabeL#1$$}
%
\newskip\footskip\footskip14pt plus 1pt minus 1pt 
\def\footnotefont{\ninepoint}\def\f@t#1{\footnotefont #1\@foot}
\def\f@@t{\baselineskip\footskip\bgroup\footnotefont\aftergroup\@foot\let\next}
\setbox\strutbox=\hbox{\vrule height9.5pt depth4.5pt width0pt}
\global\newcount\ftno \global\ftno=0
\def\foot{\global\advance\ftno by1\footnote{$^{\the\ftno}$}}
%
\newwrite\ftfile
\def\footend{\def\foot{\global\advance\ftno by1\chardef\wfile=\ftfile
$^{\the\ftno}$\ifnum\ftno=1\immediate\openout\ftfile=foots.tmp\fi%
\immediate\write\ftfile{\noexpand\smallskip%
\noexpand\item{f\the\ftno:\ }\pctsign}\findarg}%
\def\footatend{\vfill\eject\immediate\closeout\ftfile{\parindent=20pt
\centerline{\bf Footnotes}\nobreak\bigskip\input foots.tmp }}}
\def\footatend{}
%
%
\global\newcount\refno \global\refno=1
\newwrite\rfile
\def\ref{[\the\refno]\nref}
\def\nref#1{\xdef#1{[\the\refno]}\writedef{#1\leftbracket#1}%
\ifnum\refno=1\immediate\openout\rfile=refs.tmp\fi
\global\advance\refno by1\chardef\wfile=\rfile\immediate
\write\rfile{\noexpand\item{#1\ }\reflabeL{#1\hskip.31in}\pctsign}\findarg}
\def\findarg#1#{\begingroup\obeylines\newlinechar=`\^^M\pass@rg}
{\obeylines\gdef\pass@rg#1{\writ@line\relax #1^^M\hbox{}^^M}%
\gdef\writ@line#1^^M{\expandafter\toks0\expandafter{\striprel@x #1}%
\edef\next{\the\toks0}\ifx\next\em@rk\let\next=\endgroup\else\ifx\next\empty%
\else\immediate\write\wfile{\the\toks0}\fi\let\next=\writ@line\fi\next\relax}}
\def\striprel@x#1{} \def\em@rk{\hbox{}}
\def\lref{\begingroup\obeylines\lr@f}
\def\lr@f#1#2{\gdef#1{\ref#1{#2}}\endgroup\unskip}
\def\semi{;\hfil\break}
\def\addref#1{\immediate\write\rfile{\noexpand\item{}#1}} 
\def\startrefs#1{\immediate\openout\rfile=refs.tmp\refno=#1}
\def\xref{\expandafter\xr@f}\def\xr@f[#1]{#1}
\def\refs#1{\count255=1[\r@fs #1{\hbox{}}]}
\def\r@fs#1{\ifx\und@fined#1\message{reflabel \string#1 is undefined.}%
\nref#1{need to supply reference \string#1.}\fi%
\vphantom{\hphantom{#1}}\edef\next{#1}\ifx\next\em@rk\def\next{}%
\else\ifx\next#1\ifodd\count255\relax\xref#1\count255=0\fi%
\else#1\count255=1\fi\let\next=\r@fs\fi\next}
%

%
\newwrite\ffile\global\newcount\figno \global\figno=1
\def\fig{fig.~\the\figno\nfig}
\def\nfig#1{\xdef#1{fig.~\the\figno}%
\writedef{#1\leftbracket fig.\noexpand~\the\figno}%
B\ifnum\figno=1\immediate\openout\ffile=figs.tmp\fi\chardef\wfile=\ffile%
\immediate\write\ffile{\noexpand\medskip\noexpand\item{Fig.\ \the\figno. }
\reflabeL{#1\hskip.55in}\pctsign}\global\advance\figno by1\findarg}
\def\xfig{\expandafter\xf@g}\def\xf@g fig.\penalty\@M\ {}
\def\figs#1{figs.~\f@gs #1{\hbox{}}}
\def\f@gs#1{\edef\next{#1}\ifx\next\em@rk\def\next{}\else
\ifx\next#1\xfig #1\else#1\fi\let\next=\f@gs\fi\next}
\newwrite\lfile
{\escapechar-1\xdef\pctsign{\string\%}\xdef\leftbracket{\string\{}
\xdef\rightbracket{\string\}}\xdef\numbersign{\string\#}}

\def\writestop{\def\writestoppt{\immediate\write\lfile{\string\pageno%
\the\pageno\string\startrefs\leftbracket\the\refno\rightbracket%
\string\def\string\secsym\leftbracket\secsym\rightbracket%
\string\secno\the\secno\string\meqno\the\meqno}\immediate\closeout\lfile}}
\def\writestoppt{}\def\writedef#1{}
\def\seclab#1{\xdef #1{\the\secno}\writedef{#1\leftbracket#1}\wrlabeL{#1=#1}}
\def\subseclab#1{\xdef #1{\secsym\the\subsecno}%
\writedef{#1\leftbracket#1}\wrlabeL{#1=#1}}
\newwrite\tfile \def\writetoca#1{}
\def\leaderfill{\leaders\hbox to 1em{\hss.\hss}\hfill}
\def\writetoc{\immediate\openout\tfile=toc.tmp
   \def\writetoca##1{{\edef\next{\write\tfile{\noindent ##1
   \string\leaderfill {\noexpand\number\pageno} \par}}\next}}}

%
\catcode`\@=12 
%
\edef\tfontsize{\ifx\answ\bigans scaled\magstep3\else scaled\magstep4\fi}
\font\titlerm=cmr10 \tfontsize \font\titlerms=cmr7 \tfontsize
\font\titlermss=cmr5 \tfontsize \font\titlei=cmmi10 \tfontsize
\font\titleis=cmmi7 \tfontsize \font\titleiss=cmmi5 \tfontsize
\font\titlesy=cmsy10 \tfontsize \font\titlesys=cmsy7 \tfontsize
\font\titlesyss=cmsy5 \tfontsize \font\titleit=cmti10 \tfontsize
\skewchar\titlei='177 \skewchar\titleis='177 \skewchar\titleiss='177
\skewchar\titlesy='60 \skewchar\titlesys='60 \skewchar\titlesyss='60
\def\titlefont{\def\rm{\fam0\titlerm}
\textfont0=\titlerm \scriptfont0=\titlerms \scriptscriptfont0=\titlermss
\textfont1=\titlei \scriptfont1=\titleis \scriptscriptfont1=\titleiss
\textfont2=\titlesy \scriptfont2=\titlesys \scriptscriptfont2=\titlesyss
\textfont\itfam=\titleit \def\it{\fam\itfam\titleit}\rm}
 \ifx\answ\bigans\else scaled\magstep1\fi
\ifx\answ\bigans\def\abstractfont{\tenpoint}\else
\font\abssl=cmsl10 scaled \magstep1
\font\absrm=cmr10 scaled\magstep1 \font\absrms=cmr7 scaled\magstep1
\font\absrmss=cmr5 scaled\magstep1 \font\absi=cmmi10 scaled\magstep1
\font\absis=cmmi7 scaled\magstep1 \font\absiss=cmmi5 scaled\magstep1
\font\abssy=cmsy10 scaled\magstep1 \font\abssys=cmsy7 scaled\magstep1
\font\abssyss=cmsy5 scaled\magstep1 \font\absbf=cmbx10 scaled\magstep1
\skewchar\absi='177 \skewchar\absis='177 \skewchar\absiss='177
\skewchar\abssy='60 \skewchar\abssys='60 \skewchar\abssyss='60
\def\abstractfont{\def\rm{\fam0\absrm}
\textfont0=\absrm \scriptfont0=\absrms \scriptscriptfont0=\absrmss
\textfont1=\absi \scriptfont1=\absis \scriptscriptfont1=\absiss
\textfont2=\abssy \scriptfont2=\abssys \scriptscriptfont2=\abssyss
\textfont\itfam=\bigit \def\it{\fam\itfam\bigit}\def\footnotefont{\tenpoint}%
\textfont\slfam=\abssl \def\sl{\fam\slfam\abssl}%
\textfont\bffam=\absbf \def\bf{\fam\bffam\absbf}\rm}\fi
\def\tenpoint{\def\rm{\fam0\tenrm}
\textfont0=\tenrm \scriptfont0=\sevenrm \scriptscriptfont0=\fiverm
\textfont1=\teni  \scriptfont1=\seveni  \scriptscriptfont1=\fivei
\textfont2=\tensy \scriptfont2=\sevensy \scriptscriptfont2=\fivesy
\textfont\itfam=\tenit
\def\it{\fam\itfam\tenit}\def\footnotefont{\ninepoint}%
\textfont\bffam=\tenbf \def\bf{\fam\bffam\tenbf}\def\sl{\fam\slfam\tensl}\rm}
\font\ninerm=cmr9 \font\sixrm=cmr6 \font\ninei=cmmi9 \font\sixi=cmmi6
\font\ninesy=cmsy9 \font\sixsy=cmsy6 \font\ninebf=cmbx9
\font\nineit=cmti9 \font\ninesl=cmsl9 \skewchar\ninei='177
\skewchar\sixi='177 \skewchar\ninesy='60 \skewchar\sixsy='60
\def\ninepoint{\def\rm{\fam0\ninerm}
\textfont0=\ninerm \scriptfont0=\sixrm \scriptscriptfont0=\fiverm
\textfont1=\ninei \scriptfont1=\sixi \scriptscriptfont1=\fivei
\textfont2=\ninesy \scriptfont2=\sixsy \scriptscriptfont2=\fivesy
\textfont\itfam=\ninei \def\it{\fam\itfam\nineit}\def\sl{\fam\slfam\ninesl}%
\textfont\bffam=\ninebf \def\bf{\fam\bffam\ninebf}\rm}
%
%

\hyphenation{anom-aly anom-alies coun-ter-term coun-ter-terms}
\def\inv{^{\raise.15ex\hbox{${\scriptscriptstyle -}$}\kern-.05em 1}}

\def\Dsl{\,\raise.15ex\hbox{/}\mkern-13.5mu D} 
\def\dsl{\raise.15ex\hbox{/}\kern-.57em\partial}

\font\bigit=cmti10 scaled \magstep1
\def\lspace{\ifx\answ\bigans{}\else\qquad\fi}
\def\lbspace{\ifx\answ\bigans{}\else\hskip-.2in\fi} 
\def\boxeqn#1{\vcenter{\vbox{\hrule\hbox{\vrule\kern3pt\vbox{\kern3pt
	\hbox{${\displaystyle #1}$}\kern3pt}\kern3pt\vrule}\hrule}}}
\def\mbox#1#2{\vcenter{\hrule \hbox{\vrule height#2in
		\kern#1in \vrule} \hrule}}  
%

 \def\CR{{\cal R}}  
\def\e#1{{\rm e}^{^{\textstyle#1}}}

\def\darr#1{\raise1.5ex\hbox{$\leftrightarrow$}\mkern-16.5mu #1}

\def\roughly#1{\raise.3ex\hbox{$#1$\kern-.75em\lower1ex\hbox{$\sim$}}}


\def\mm{{=}}
\def\pp{{+\!\!\!+}}
\def\ll{\char 32l}
\def\o{\omega}
\def\O{\Omega}
\def\s{\sigma}
\def\l{\lambda}
\def\e{{\rm e}}
\def\mnbox#1#2{\vcenter{\hrule \hbox{\vrule height#2in
                \kern#1in \vrule} \hrule}}  
\def\sq{\,\raise.5pt\hbox{$\mnbox{.09}{.09}$}\,}
\def\sqb{\,\raise.5pt\hbox{$\overline{\mnbox{.09}{.09}}$}\,}
\def\sqr#1#2{{\vcenter{\vbox{\hrule height.#2pt
     \hbox{\vrule width.#2pt height#1pt \kern#1pt
           \vrule width.#2pt}
       \hrule height.#2pt}}}}

\Title{\vbox{\baselineskip12pt\hbox{USITP-93-04}
\hbox{hep-th/9303109}
}}
{{\vbox{\centerline{ Induced chiral  supergravities in 2D}}}}

\centerline{
Fiorenzo Bastianelli\footnote{$^\dagger$}{e-mail: fiorenzo@vana.physto.se}
 and
Ulf Lindstr\"om\footnote{$^{\dagger \dagger}$}{e-mail: ul@vana.physto.se}}
\bigskip\centerline
{\it  Institute for Theoretical Physics}
\centerline {\it  Stockholm University}
\centerline {\it Box 6730}
\centerline {\it S-113 85 Stockholm, Sweden}
\vskip .5in

\noindent
We analyze actions for 2D supergravities  induced by chiral conformal
supermatter. The latter
may be thought as described  at the classical level by
superspace actions
invariant under super-reparametrization, super-Weyl  and super-Lorentz
transformations. Upon quantization various anomalies appear which
characterize the non-trivial induced actions for the
supergravitational sector. We derive these induced actions
using a chiral boson to represent the chiral inducing matter.
We show that they can be defined in a super-reparametrization invariant
way, but with super-Weyl and super-Lorentz anomalies.
We consider the case of $(1,0)$ and $(1,1)$ supergravities
by working in their respective superspace formulations and
investigate  their quantization in the conformal gauge.
The  actions we consider arise naturally in off-critical
 heterotic and spinning  strings.
In the conformal gauge, they correspond to chiral extensions of the
super-Liouville theory.

\vskip .8cm

\Date{3/93. \ \ \  Revised version 6/93.}
\eject

\newsec{Introduction}

In the Polyakov approach to string theory
\ref\Pol{A. Polyakov, Phys. Lett. B103 (1983) 207,  ibid.   211. },
one has to consider  the quantization of two dimensional matter
coupled  to gravity. The gravitational sector can be
 described by the vielbein field $e_\mu{}^a$ and
has no propagating degrees of freedom in the critical dimensions.
In fact, it can be gauge fixed
using reparametrizations, Weyl rescaling and Lorentz rotations
to a given background value, up to some global configurations
described by the moduli.  Thus,
there is no local dynamics for two dimensional gravity
in the target spacetime critical dimensions, since
in such dimensions all of the local symmetries used to
gauge away the vielbein are free of anomalies.
In the bosonic string, one can generalize the concept of
dimension by the pair of numbers given by the values
of the central charges, $c$
and $\bar c$, of the two copies
of the Virasoro algebra generated by the matter stress tensor.
The critical dimensions correspond in this language to $c=\bar c=26$,
most easily realized using
26 scalar fields which are
interpreted  as the coordinates of the string moving in a
26 dimensional flat spacetime.
Evidently, a similar point of view can be taken also
for the various supersymmetric extensions of the
bosonic string theory, where one replaces the two
Virasoro algebras by the appropriate constraint algebra defining the
given string theory, e.g.  super-Virasoro$\otimes$Virasoro
for the heterotic string or super-Virasoro$\otimes$super-Virasoro
for the spinning string.
By a string \lq\lq off the critical dimensions\rq\rq, one then refers
to string models  in which the matter central charges
do not saturate the critical bound. For example, in the bosonic case
for $c=\bar c\neq 26$  an anomaly appears in the Weyl symmetry \Pol,
and the conformal factor of the two dimensional vielbein
becomes dynamical.
This can be seen computing the induced gravitational
 action, obtained by integrating out
the matter fields in the path integral.
The resulting action  $I \sim c R{ \sq }^{-1} R $ is not Weyl invariant
and the reparametrization ghosts,  which contribute
$I_{gh} \sim -26 R{\sq }^{-1} R $, are not sufficient
to cancel the Weyl  anomaly.
Therefore, one is led to study the  quantization of the action
$ c R{\sq}^{-1} R $, which reduces to
 the Liouville model in the conformal
gauge,  to be able to
complete the string quantization   off the critical dimensions.
Much progress has been achieved in such a task for $c=\bar c \le 1$. It
begun with the works of  Polyakov
\ref\P{A. Polyakov, Mod. Phys. Lett. A2 (1987) 893.}
and Knizhnik, Polyakov and  Zamolodchikov
\ref\KPZ{ V. Knizhnik, A. Polyakov and A.B. Zamolodchikov,
 Mod. Phys. Lett. A3 (1988) 819.},
who used a light-cone gauge for the metric, and followed by  the works of
David
\ref\Dav{ F. David, Mod. Phys. Lett. A3 (1988) 1651.} and
Distler and Kawai
\ref\DK{ J. Distler and H. Kawai, Nucl. Phys. B231 (1989) 509.},
who instead used
the more natural conformal gauge.
In these works, exact results in the form of
gauge invariant critical exponents
were obtained.
Eventually, dynamical triangulations and matrix models
in the double scaling limit were
successfully used
to improve on these results
\ref\mat{ M.R.  Douglas and S.H. Shenker, Nucl. Phys. B335 (1990) 635\semi
E. Br\'ezin and V.A. Kazakov, Phys. Lett. 236B (1990) 144\semi
D. Gross and A.A. Migdal, Phys. Rev. Lett. 64 (1990) 127, Nucl Phys. B340
(1990) 333.},
 at least for their relevance to
string theory, allowing to sum up the  perturbative expansion
in the topology of the worldsheet.
Efforts are now  being made to understand how to overcome the $c=1$
barrier, see for example  \ref\Rag{
E. Br\'ezin and S. Hikami, Phys. Lett. B283 (1992) 203\semi
L. Alvarez Gaum\'e,
J.L.F. Barb\'on and C. Crnkovi\'c, Nucl. Phys. B394 (1993) 383\semi
H. Kawai and R. Nakayama, Phys. Lett. B306 (1993) 224.}.
However, there is a
more general way to interpret strings off the
critical dimensions. Namely, one can imagine
the  situation in which  the matter central charges $c$ and $\bar c$
are different from each other and both different from 26.
This happens only if the matter fields living on the worldsheet are chiral.
It is obviously an interesting and phenomenologically promising
situation, since chiral structures
on the worldsheet will typically induce chiral properties in spacetime.
Examples are easily constructed
using two dimensional Majorana-Weyl fermions, which have
$ c={1\over 2} $ and $\bar c =0$ (or vice-versa for the opposite chirality).
New kind of
anomalies arise in such a  situation, namely the gravitational anomalies
\ref\AG{L. Alvarez-Gaum\'e and E. Witten, Nucl. Phys. B234 (1984) 269.}.
However, it is known that one can shift these anomalies into the
Weyl and Lorentz sectors by using local counterterms \ref\BZ{W.A. Bardeen and
B. Zumino, Nucl. Phys. B234 (1984) 421\semi
L. Bonora, P. Pasti and M. Tonin, J. Math. Phys 27 (1986) 2259.}.
It is in this latter form that the full effective action
for the case of a Weyl fermion was first found
by Leutweyler
\ref\Le{H. Leutweyler, Phys. Lett.  B153 (1985) 65.}.
It can be parametrized  as follows
\eqn\effac{ \eqalign{ &\e^{- I[e_\mu{}^a]} = \int ({\cal D} X)_e \
\e^{-S[X,e_\mu{}^{a}]}
\cr
&I[e_\mu{}^{a}] = {1\over 24 \pi} \int d^2 x e
\biggl ( c R_1 {1\over \sq} R_1 +
\bar c R_2 {1\over \sq} R_2
+ 2a \o_\pp \o_= \biggr ),\cr}}
where $X$ represents a generic chiral conformal system
with central charges $c$ and $\bar c$ used to induce
the gravitational action.
The coupling $a$  multiplies a local term
and is not fixed by the requirement of diffeomorphism invariance,
 $R_1$ and $R_2$ are Lorentz non-invariant
chiral pieces of the curvature
scalar  constructed out of the vielbein field  and $\o_a$
are the components of the spin connection
(the flat index $a$ takes the values $ (\pp,=)$,
see appendix A for notation).
Quantization of this induced action for chiral gravity (or, equivalently,
chirally induced action for gravity!) has been recently
investigated by Oz,  Pawe{\ll}czyk and  Yankielowicz
\ref\Yaron{Y. Oz,  J. Pawe{\ll}czyk and S. Yankielowicz,
              Nucl. Phys. B363 (1991) 555.}
in a light-cone gauge for the metric, and  by Myers and  Periwal
\ref\Myers{R. Myers and V. Periwal, Nucl. Phys. B395 (1993) 119;
{\it Chiral non-critical strings},
 preprint Mc-Gill/92-47 and hep-th/9210082. }
in the conformal gauge\footnote
{$^\dagger$}{A mismatch between the results of refs. \Yaron\ and \Myers\
is properly understood
in \ref\fio{F. Bastianelli, Mod. Phys. Lett. A7 (1992) 3777.},
where it is checked that the stress tensor for the Lorentz
field  used in \Yaron\ is  the one arising from \effac,
but with a fixed value of $a$. This value can in fact be left
arbitrary.}.

In the present paper, we consider the case of off-critical
heterotic and spinning string by first deriving the chirally
induced action for the corresponding supergravities.
Since the constraint algebra is made out of a left $N=1$
super-Virasoro plus a right Virasoro algebra for the heterotic string,
 and  of a   $N=1$ super-Virasoro algebra in each chiral
sector for the spinning string, the
supergravities  to consider are  $(1,0)$ and $(1,1)$ respectively.
We identify the chirally induced actions with a simple trick.
We use  chiral bosons to represent a general chiral system.
We evaluate the  gaussian path integral by square
completion and from the knowledge of the
action induced by  a free scalar field we obtain the  required result.
The chiral boson  we use is the one recently introduced in ref.
\ref\fiore{ F. Bastianelli, Phys. Lett. B277 (1992) 464.}, and  consists
of a scalar field with peculiar couplings to the
background geometry. After having  obtained the induced  actions,
we proceed  as in David, Distler and Kawai \Dav \DK\
to analyze  their quantization in the conformal gauge,
where they reduce to local actions for the Weyl and Lorentz modes
of the vielbein. We do  not identify any new exact critical exponents,
partly because we only give a local description of the model and do not
address topological issues.

The structure of the paper is as follows.
In sec. 2 we review the case of chiral gravity, explaining our strategy
in a simpler context and in a way that generalizes
to the supersymmetric case. In sec. 3 we discuss the off-critical
heterotic string, presenting the chiral induced action
for  $(1,0)$ supergravity and quantizing it in the conformal gauge using
free fields. We check the independence on the gauge fixing choice
by verifying that the the Lorentz and Weyl
anomalies cancel. This is a necessary requirement for the BRST invariance
of the quantum theory.
In sec. 4 we repeat our analysis for the $(1,1)$
case, i.e. for the spinning string. From one point of view,
this analysis is simpler than that for the heterotic string,
since the (1,1) superspace is intrinsically non-chiral
and much of the derivations are in close parallel with the bosonic case.
Eventually, in sec. 5 we present our conclusions and an outlook.
We explain our conventions and notations in the appendices,
where we review the various superspaces and list few useful
formulae employed in the main text.


\newsec{Review of induced chiral gravity}

The easiest way to obtain the gravitational action induced
by chiral matter in 2d, eq. \effac, is to use a chiral
boson with central charges
$c$ and $\bar c$ to represent the  inducing matter system.
The chiral boson we have in mind was
introduced by one of us in ref. \fiore\
and is described by the action
\eqn\cb{ S[X,e_\mu{}^a] = {1\over {2 \pi}} \int d^2x e ( \nabla_\pp X
\nabla_= X + \beta_1 R_1 X + \beta_2 R_2 X), }
where $R_1$ and $R_2$ are chiral halves of the curvature scalar
$R=R_1+R_2$.  The precise definition of $R_1$ and $R_2$ is
to be found in appendix A,
to which we refer also for further details about our notation.
The coupling to the chiral curvature scalars induce
improvement terms in the stress tensor, which is traceless and conserved when
evaluated in flat space, and generates
two copies of a Virasoro algebra with central charges
$c= 1+ 3 \beta_1^2$ and  $\bar c = 1+ 3 \beta_2^2$.
This is seen as follows. From \cb\ we compute the stress tensor,
defined as
\eqn\dst{T_{ab} = {{2 \pi} \over e} {\delta S \over {\delta e_\mu{}^a}}
e_\mu{}_b. }
We evaluate it on-shell and in flat space, obtaining the following
non-vanishing components
\eqn\nvc{\eqalign {
T_{\pp\pp} &= - {1\over 2} \partial_\pp X \partial_\pp X+
{\beta_1 \over 2} \partial_\pp^2
X  \cr
T_{==} &=-{1\over 2}\partial_=X \partial_=X
 +{\beta_2 \over 2} \partial_=^2
X .\cr}}
Using now the propagator derived from the flat space limit of \cb,
\eqn\dprop{\langle X(x) X(y) \rangle =-\log (\mu^2 | x-y|^2)}
with $\mu$ an infrared cut-off, it is immediate to obtain
\eqn\ccbar{c= 1+ 3 \beta_1^2, \ \ \ \ \ \bar c = 1+ 3 \beta_2^2}
for the central charges $c $ and $\bar c$
of the Virasoro algebras generated by
the operator product expansions of
$T_{\pp\pp}$ with itself and $ T_{==}$ with itself, respectively.
This confirms that the system described by
the action \cb\ has chiral properties.

Let's now consider the  gravitational
action induced by chiral
systems. In order to derive it,
we write down the path integral for the chiral boson
\eqn\picb{ \e^{- I[e_\mu{}^a]} = \int ({\cal D} X)_e \
\e^{-S[X,e_\mu{}^{a}]} }
where the translational invariant
measure $({\cal D} X)_e$ is implicitly defined using
the following ultralocal and reparametrization invariant
 metric on field space (see e.g. refs.
\ref\JP{ J. Polchinski, Comm. Math. Phys. 104 (1986) 37.},
\DK\ and
references therein)
\eqn\measure{ || \delta X ||^2_e =
 \int d^2x e  (\delta X)^2 , \ \ \ \ \int ({\cal D} \delta X)_e \
\e^{- || \delta X ||^2_e }= 1  .}
We compute this path integral  by completing
squares and shifting the field
 $ X\rightarrow X + {1\over 2} {\sq}^{-1} ( \beta_1 R_1
+\beta_2 R_2)$, so that it reduces to the path integral of
a free scalar without background charges.
This way we  obtain the following induced action
\eqn\ind{ I [e_\mu{}^a] = {1\over 24 \pi} \int d^2x e
\biggl (  R {1\over \sq} R  +
3 (\beta_1 R_1 + \beta_2 R_2 ){1 \over \sq} (
\beta_1 R_1 + \beta_2 R_2)
\biggr ),}
where the first term on the right hand side is due to
the well-known contribution of a
 free boson (which has $c=\bar c=1$) and the
second term is due to square completion.
Cross terms between $R_1$ and $R_2$
are local (up to boundary terms discarded in our local analysis)
\eqn\lt{ \int d^2x e
R_1 {1\over \sq} R_2  =
\int d^2x e \omega_\pp \omega_= }
and eq. \ind\ can be written as follows
\eqn\acn{ I[e_\mu{}^{a}] =
I[e_\mu{}^{a}; c,\bar c,a] \equiv {1\over 24 \pi} \int d^2 x e
\biggl ( c R_1 {1\over \sq} R_1 +
\bar c R_2 {1\over \sq} R_2
+ 2a \o_\pp \o_= \biggr ) ,}
with $c$ and $\bar c$ given by eq. \ccbar.
The parameter $a$ is ambiguous and whereas the above computation gives
$a=1+ 3\beta_1 \beta_2$,  its value is generically
 related to the specific regularization
procedure adopted for computing the induced action.
It multiplies a local term, and it can be changed at will by
adding a local counterterm of the same form to the effective action.
It is not fixed by requiring
general coordinate invariance and 2d rigid Lorentz invariance.
Note that for $c=\bar c$, one can
recover  the local Lorentz invariance by
choosing $a=c$, thus  obtaining the non-chiral
action for gravity
\eqn\ag{ I_0 [e_\mu{}^a; c] = {c\over 24 \pi} \int d^2x e
  R {1\over \sq} R .}

We consider now the action in
\acn\ as the gravitational action to be quantized in order to
investigate properties of off-critical chiral strings.
We again point out
that such an action is manifestly reparametrization invariant
because it is built from manifestly invariant
objects (scalars), but contains Lorentz and Weyl anomalies.
It gives dynamics to the Lorentz and Weyl modes of the vielbein.
To see this explicitly, we choose
the conformal gauge for the diffeomorphism group
\eqn\cong{ e_\mu{}^\pp = \exp (\sigma-i\lambda) \hat  e_\mu{}^\pp,\ \ \
e_\mu{}^= = \exp (\sigma +i\lambda) \hat  e_\mu{}^= ,}
where $\hat e_\mu{}^a$ is a given background vielbein, and  obtain
from \acn\ the following Lorentz-Weyl action
\eqn\wla{\eqalign{ S[\lambda,\sigma,\hat e_\mu{}^a]
&\equiv I[e_\mu{}^a] - I[\hat e_\mu{}^a]  \cr &=
{1\over 24 \pi} \int d^2x \hat e
\bigl [\sigma \hat  {\sq} \sigma ( c+ \bar c + 2a)
       + \lambda \hat  {\sq} \lambda ( 2a  -c - \bar c ) +
       \lambda \hat  {\sq} \sigma  2i ( \bar c - c )   \cr & \ \ \ \
  - 2  \hat R_1 \sigma (c+a)   - 2  \hat R_2 \sigma (\bar c + a)
- 2i\hat R_1 \lambda (a-c) -2 i \hat R_2 \lambda (\bar c - a) \bigr ].
\cr}} This is a generalization of the Liouville action
(with the cosmological constant set to zero)
induced by off-critical strings. Note that
we have written our  equations with an euclidean signature
for the worldsheet,
where the Lorentz  group is  the compact group $ U(1)$
and the $\lambda$
field is compactified.
In a minkowskian signature one should Wick rotate to
$\varphi=i\lambda$ and forget about the compactness.
These comments should be
kept in mind, for example, when discussing unitarity of the theory
or its global properties.

We are now going to analyze the
quantization of \wla. For simplicity we proceed
in a stepwise fashion, first considering  the
quantization of the Lorentz field
$\lambda$ and then the quantization of the
Weyl field $\sigma$. We will use methods similar to those employed
by David, Distler and Kawai, i.e. we will
 disregard the cosmological constant, which we have
already set to zero from the beginning, and employ free field quantization.
Therefore, we start anew by making explicit the dependence
of the induced action on the Lorentz
field only.
We set $ e_\mu{}^\pp = \exp (-i\lambda) \tilde  e_\mu{}^\pp$
and $ e_\mu{}^= = \exp (i\lambda) \tilde  e_\mu{}^= $
 as a partial gauge choice,
and obtain from  \acn\ the following
action for the Lorentz field  (we just have to
set $\sigma=0$ in \wla\ and replace hats with tildes)
\eqn\la{\eqalign{
S_{Lor}[\lambda, \tilde e_\mu{}^a]
&\equiv I[e_\mu{}^a] - I[\tilde e_\mu{}^a]
 \cr &=
{1\over 24 \pi} \int d^2x \tilde e
\bigl [ \lambda \tilde  {\sq} \lambda ( 2a  -c - \bar c )
- 2i \tilde R_1 \lambda (a-c) - 2 i  \tilde R_2 \lambda (\bar c - a)
    \bigr ]. \cr}}
Of course, the background $\tilde e_\mu{}^a$ now contains the Weyl
field $\sigma$: $\tilde  e_\mu{}^a =  \exp (\sigma)\hat  e_\mu{}^a$.
{}From this expression it looks
that $\lambda$ behaves like a chiral boson
with background charges, as in \cb.
To be sure we have to check that no
surprises arise from the correct path integral measure
that must be used in quantizing $\lambda$.
Such a measure is the one induced by the reparametrization
invariant metric on the space of worldsheet vielbeins and
it is  determined by
\eqn\piml{ || \delta \lambda ||_{ e}^2 = \int d^2x
 e (\delta \lambda)^2
\ \  \to ({\cal D} \lambda)_{ e}.}
Of course, we can substitute  $\tilde e$ for $e$, since the
determinant of the vielbein is Lorentz invariant, implying that
$({\cal D} \lambda)_{ e} = ({\cal D} \lambda)_{\tilde e}$.
We see that the functional measure  for
the Lorentz field coincides with the usual translational invariant
measure used to quantize free scalars, as in \picb, and we
conclude that $\lambda$ can really be  treated as a chiral boson.
It is then immediate to write down
the components of the stress tensor and their central charges
(just by using the previously derived formulas for the chiral boson)
\eqn\stress{\eqalign {
T_{\pp\pp} &= - {1\over 2} (\partial_\pp \lambda)^2 +
{\beta_1 \over 2} \partial_\pp^2
\lambda,  \ \
\beta_1 = i {{c-a}\over 6}
\biggl ( {12 \over { c +\bar c -2a}} \biggr)^{ 1\over 2},\ \ \
 c_{Lor}= 1 - {(c-a)^2 \over {c +\bar c - 2a}} ,\cr
T_{==} &=-{1\over 2}(\partial_=\lambda)^2 +{\beta_2 \over 2}
\partial_=^2 \lambda, \ \ \
\beta_2 = i {{ a-\bar c}\over 6}
\biggl ( {12 \over { c +\bar c -2a}} \biggr)^{ 1\over 2},\ \ \
\bar c_{Lor}= 1 - {(\bar c-a)^2 \over {c +\bar c - 2a}} ,\cr}}
where we have rescaled the Lorentz field $\lambda \rightarrow
\lambda \bigl ( {12 \over { c +\bar c -2a}} \bigr)^{ 1\over 2}$
to get a standard normalization for the propagator
\eqn\prop{\langle \lambda(x) \lambda(y) \rangle
=-\log (\mu^2 | x-y|^2).}
In doing so we have assumed $c+\bar c-2a>0$.
The same holds also for $c+\bar c-2a<0$, but in this case we have to
remember that with the field redefinition we have
automatically performed a Wick rotation
on $\lambda$.
The Lorentz field has the effect of leveling up the chiral mismatch
between the matter central charges $c$ and $\bar c$, namely
\eqn\lev{ c+ c_{Lor} = \bar c+\bar c_{Lor} =
1 + { {c \bar c - a^2 }\over {c +\bar c - 2a}}.}
Note that $a$ is left arbitrary.
We can derive once more these results by path integrating
the Lorentz field with the measure in \piml. Completing the
 squares  and using  the translational invariance of the measure
we obtain
\eqn\cunopi{
 \eqalign{ &\e^{- I_{Lor}[\tilde e_\mu{}^a]} = \int ({\cal D}
\lambda)_{\tilde e} \
\e^{-S_{Lor}[\lambda ,\tilde e_\mu{}^{a}]} \cr
&I_{Lor}[\tilde e_\mu{}^{a}] = I[\tilde e_\mu{}^{a}; c_{Lor},\bar
c_{Lor}, a_{Lor}] \cr}}
 where the functional on the right hand
side of the second equation was defined in \acn, $  c_{Lor}$
and $\bar c_{Lor}$ are as in \stress, and
\eqn\alor{  a_{Lor} = 1+ {(c-a)(\bar c -a) \over {c +\bar c - 2a}}.}
In principle the value of $a_{Lor}$ could be changed by adding a local
counterterm to $I_{Lor}$, but this is not necessary since we can check that
$ a +a_{Lor} =  c +c_{Lor}$. It is the correct value
which secures  background Lorentz invariance
\eqn\comb{ I[\tilde  e_\mu{}^a] +
I_{Lor}[\tilde  e_\mu{}^a]
= I_0[\tilde  e_\mu{}^a, c + c_{Lor}].}
The functional on the right hand side of this equation was defined in \ag.
Since $a$ is still a free parameter, one could fix it by requiring
$ c +c_{Lor} = 26$, implying that the Weyl invariance
is recovered once
the reparametrization ghosts are introduced.
We will avoid such a fine tuning, so that we are still left
 to quantize the Weyl mode $\sigma$.
This can be done by repeating the
analysis of David, Distler and Kawai, and goes as follows.
We complete the specification of the conformal gauge by setting
$\tilde  e_\mu{}^a =  \exp (\sigma)\hat  e_\mu{}^a$
and take into account the well-known contribution
of the reparametrization ghosts
to the induced gravitational action
\eqn\repgh{ I_{gh}[\tilde  e_\mu{}^a] = I_0[\tilde  e_\mu{}^a; -26].}
The reparametrization ghosts are non-chiral
and this fact has allowed
us to postpone their inclusion up
to now.  Introducing them at an earlier stage would not have affected
the previous analysis.
The combined matter, Lorentz and ghost fields
induce the non-chiral action
$I_0[\tilde  e_\mu{}^a, c + c_{Lor}-26]$
which in turn produces the following Liouville action for the Weyl field
\eqn\weyla{\eqalign{
S_{Liou}[\sigma, \hat e_\mu{}^a]
&\equiv I_0[\tilde e_\mu{}^a; c+c_{Lor} -26] - I_0[\hat e_\mu{}^a;
c+c_{Lor} -26]
 \cr &= ({c+c_{Lor} -26})
{1\over 6 \pi} \int d^2x \hat e
\bigl [ \sigma \hat  {\sq} \sigma - \hat R \sigma
    \bigr ]. \cr}}
This Liouville action should be quantized with the measure
induced by the distance function
\eqn\pimwuno{ || \delta \sigma ||_{ e}^2 =
\int d^2x  e (\delta \sigma)^2=
\int d^2x \hat  e \e^\sigma (\delta \sigma)^2
\ \  \to ({\cal D} \sigma)_{ e}}
which is derived from the reparametrization invariant metric on the space of
worldsheet vielbeins. The problem with this measure
is that it is not invariant under shifts of the field $\sigma$  and it is not
known how to use it to quantize the theory. The best one can do
is to  quantize the Weyl field using the shift invariant
measure obtained from
\eqn\pimwdue{ || \delta \sigma ||_{\hat e}^2 = \int d^2x
\hat e (\delta \sigma)^2
\ \  \to ({\cal D} \sigma)_{\hat e}}
and including a jacobian $J$ which relates \pimwuno\ to \pimwdue\ \DK\
\eqn\jac { ({\cal D} \sigma)_{e}= J
({\cal D} \sigma)_{\hat e}.}
This jacobian is formally given by
\eqn\jac{ J= {\rm det}\ \e^\sigma }
and was computed in
\ref\MM{N.E. Mavromatos and J.L. Miramontes, Mod. Phys. Lett. A4 (1989)
1847\semi
E. D'Hoker and P.S. Kurzepa, Mod. Phys. Lett. A5 (1990) 1441.}
using an heat kernel regularization.
The heat kernel used was the one corresponding to the scalar laplacian,
which is the second functional derivative of the Liouville action
\weyla\ (i.e. it corresponds to the kinetic operator for the Weyl field).
This means, in practice, that
to compute the jacobian $J$ one has to compute the gravitational
action induced by  a single scalar field. In fact,
the two computations are similar and the only difference
is a normalization factor easily taken into account.
The result is thus given by
\eqn\result{J = \exp \bigr (- (I_0[ \tilde e_\mu^a; 1]- I_0[ \hat e_\mu^a;
1])
\bigl ).}
The subtraction of the constant $ I_0[ \hat e_\mu^a; 1]$ insures
that $J=1$ for $\sigma =0$.
Including the contribution of this jacobian in \weyla\
brings us to consider the following path integral for the
quantization of the Weyl mode
\eqn\pippo{\eqalign{ &\e^{- I_{Liou}[\hat e_\mu{}^a]
} = \int ({\cal D}
\sigma)_{\hat e} \
\e^{-S_{Liou}[\sigma ,\hat e_\mu{}^{a}]} \cr
&S_{Liou}[\sigma,\hat e_\mu{}^{a}] = ({c+c_{Lor} -25})
{1\over 6 \pi} \int d^2x \hat e
\bigl [ \sigma \hat  {\sq} \sigma - \hat R \sigma
    \bigr ]\cr}}
which gives
\eqn\pippodue{\eqalign{
&I_{Liou}[\hat e_\mu{}^a] = I_0[\hat e_\mu{}^a; c_{Liou}]\cr
& c_{Liou}= 26-c-c_{Lor}.\cr}}
It implies that with the inclusion of the Weyl field
the total central charge vanish in each chiral sector
\eqn\zero{ \eqalign{& c_{tot}= c + c_{Lor} - 26 + c_{Liou} =0 \cr
&\bar c_{tot}= \bar c + \bar c_{Lor} - 26 +
\bar c_{Liou} =0, \cr}}
or, equivalently,
\eqn\epp{ I[\hat e_\mu{}^{a}] + I_{Lor}[\hat e_\mu{}^{a}] +
I_{gh}[\hat e_\mu{}^{a}] + I_{Liou}[\hat e_\mu{}^{a}] =0.}
It is tempting to identify $ c'\equiv c + c_{Lor}$
as the central charge $c$ appearing  in the formulas
for non-chiral gravity \KPZ \Dav \DK, carrying
over all of the corresponding expressions for critical indices and observing
that the barrier  $c=1$ is replaced by $c'=1$.
It implies that in term of our $c$, the barrier is $a$-dependent
and can be avoided even for $c>1$ by properly choosing  $a$.
However, this conclusion seems too na{\"\i}ve to us , and while it is partly
confirmed in \Myers,  we believe that
more analysis is needed to understand: $i)$ the topological
issues connected to superselection rules for the Lorentz field
found by Myers and Periwal, expecially from the point of view of path
integrals,  $ii)$
unitarity of the theory, since the imaginary
coupling and the possibility of getting $ c_{Lor}$ and $ c'$ negative
by choosing $a$ looks suspicious.
We stress once more that the quantization of the
gravitational sector in the conformal gauge
has resulted in a Lorentz-Weyl theory
with central charges $c_{grav} \equiv c_{Lor} + c_{Liou}$ and
$\bar c_{grav} \equiv \bar c_{Lor} + \bar c_{Liou}$ obeying
$c_{grav}= 26 - c$ and $\bar c_{grav}=26 - \bar c $.
We have shown this  by quantizing the Lorentz field first
and  the Weyl field afterwards.
The vanishing of the total central charges is needed for
the  consistent quantization of the  gravitational sector,
since background Lorentz
and Weyl symmetries correspond to a change of the gauge slice
required in fixing the
reparametrization invariance.
Of course, the final result on the vanishing of the total central charges
is independent of any stepwise analysis \Myers.
In the sequel we will
develop a similar analysis for the $(1,0)$ and $(1,1)$ supersymmetric
cases and will not dwell further on the interesting but difficult
topological issues mentioned above, nor on the question of unitarity.

To conclude this section, we summarize the strategy just presented,
since we will follow it closely in the announced supersymmetric
extensions. First, we construct a chiral boson to represent
a conformal system with arbitrary central charges $c$ and $\bar c$.
Then, we use it to induce the gravitational action containing the
expected  Weyl and Lorentz anomalies. This is the action which should
be quantized in order to describe  off-critical string theories.
Then, we proceed to discuss some aspects of this quantization
using the free field approach pioneered by David, Distler and Kawai.
We check that at the quantum level the Lorentz mode behaves as a chiral
boson. This allows us to directly apply formulae already in our hands.
A similar analysis then applies to the Weyl mode, which
 is non-chiral in the $(0,0)$ and $(1,1)$ cases.
In the $(1,0)$ case, an unavoidable chiral structure for the Weyl mode
is present, as dictated by the  chiral structure of the superspace
itself.  Looking at the
total central charges we check that the background Lorentz and Weyl
symmetries hold after quantization, a necessary requirement for
gauge independence.


\newsec{Induced $\bf (1,0)$ chiral supergravity}

To identify the induced action for the heterotic string off-critical
dimensions, we are going to parallel the previous analysis in the
$(1,0)$ superspace, reviewed for convenience in appendix B.
To start with, we describe a chiral boson in superspace.
Of course, the usual free scalar superfield $X$ with action
\eqn\scs{ S_{f}[X, E_M{}^A]=
{1\over 2\pi} \int d^3Z e_+  \nabla_+ X \nabla_= X }
is already chiral because of the chiral nature of $(1,0)$ superspace.
It contains a scalar field plus a left moving Majorana-Weyl fermion.
The corresponding stress tensor generates a left superconformal algebra
with $c={3\over2}$ and a right conformal algebra with $\bar c=1$.
To achieve arbitrary central charges we consider the coupling
of the field $X$ to the
background curvatures $R_1^+$ and $R_2^+$ described in appendix B,
so that the stress tensor will acquire improvement terms
\eqn\chi{ S[X, E_M{}^A]= {1\over 2\pi} \int d^3Z e_+
(\nabla_+ X \nabla_= X + \beta_1 R_1^+ X + \beta_2 R_2^+ X).}
The super-stress tensor $T_{-A}{}^B$ is defined by
\eqn\csst{\delta S = - {1\over \pi}
\int d^3Z e_+ T_{-A}{}^B H_B{}^A (-1)^A}
with $H_A{}^B= \delta E_A{}^M E_M{}^B$ running over the set of
independent variations $(H_+{}^A, H_={}^=, H_={}^\pp)$
(the variations $H_A{}^B$  are not all independent  because
the constraints defining the  superspace must be satisfied).
Employing the list of dependent variations reported in appendix B,
we obtain the following non-vanishing components of the stress tensor
evaluated on-shell and in flat superspace
\eqn\csstress{\eqalign{ T &\equiv T_{-\pp}{}^=
= -{1\over 2} D_+ X \partial_\pp X - {1 \over 2} \beta_1
 \partial_\pp D_+ X, \cr
\bar T &\equiv T_{-\mm}{}^+
= -{1\over 2} \partial_\mm X  \partial_\mm X
- {1 \over 2}  \beta_2 \partial_\mm^2  X. \cr}}
These components of the stress tensor generate
through the operator product expansion a
superconformal algebra with $c={3\over 2} + 3\beta_1^2$ and a
conformal algebra with $\bar c= 1 + 3\beta_2^2$.
To verify this, one just need to use the propagator which follows
from the flat space limit of \scs
\eqn\csprop{ \langle X(Z_1) X(Z_2) \rangle = - \log
\bigl (\mu^2 z_{12}^\pp ( z_1^= - z_{2}^= )\bigr ), }
where $z_{12}^\pp= z_1^\pp - z_2^\pp - \theta_1^+ \theta_2^+$,
and perform the necessary Wick contractions.

We now compute the $(1,0)$ supergravitational induced action.
We first need to recall that for $c=\bar c$ the super-Weyl
anomalous action was obtained in \ref\GG{S.J. Gates, M. Grisaru, L.
Mezincescu and P. Townsend, Nucl. Phys. B286 (1987) 1\semi
R. Brooks, F. Muhammad and S.J. Gates, Nucl.Phys. B268 (1986) 599.}
and reads
\eqn\cssnca{ I_0[E_M{}^A;c] =  {c\over{ 24 \pi}} \int d^3Z e_+
 R^+ {1\over { \sq +{1\over 4}  R^+\nabla_+}} \nabla_+ R^+ }
where $ \sq = {1\over 2} ( \nabla_\pp \nabla_\mm + \nabla_\mm \nabla_\pp)$.
It can be cast in a more convenient form which is
reminiscent of the non-chiral bosonic induced action
\eqn\cssncad{ I_0[E_M{}^A;c]
 =  {c\over 24 \pi} \int d^3Z e_+
 R^+ {1\over \nabla_+\nabla_=} R^+. }
Using this piece of information and the chiral boson
constructed above, we are able to obtain the general
$(1,0)$ induced action with super-Weyl and super-Lorentz anomalies.
It is given by
\eqn\cssea{ I[E_M{}^A] =
I[E_M{}^A; c,\bar c,a] \equiv   {1\over 24 \pi} \int d^3Z e_+
\biggl ( c R_1^+ {1\over \nabla_+ \nabla_=} R_1^+ +
\bar c R_2^+ {1\over \nabla_+  \nabla_=} R_2^+
+ 2a \O_+ \O_= \biggr ) }
where $c$ and $\bar c$ are the central charges characterizing
the two chiral symmetry algebras of the inducing matter, and
$a$ is a coupling left unfixed by the requirement of reparametrization
invariance (it can be changed at will by adding a
local counterterm of the form
$ \O_+ \O_= $ to the induced action).
This result can be obtained by the following considerations.
 A system with
 $c=\bar c= {3\over2}$ can be constructed  by setting $\beta_1=0$
and $\beta_2 = {1\over{\sqrt 6}}$ in \chi.  It induces \cssncad\
with $c={3\over2}$ once we use the path integral measure derived
from the  super-reparametrization invariant norm
\eqn\duedis{ || \delta X||_{ E}^2 =
\int d^3Z  e_+ (\delta  X)^2
\rightarrow ({\cal D} X)_{ E} }
joined with the requirement of fixing local ambiguities
in the induced action by imposing
Lorentz invariance. This is so because
the path integral measure is invariant under super-reparametrizations,
leaving only the possibility of Lorentz and Weyl anomalies. The Lorentz
anomalies are eliminated by the requirement of preserving the corresponding
Ward identities, which is possible for $c=\bar c$, leaving only Weyl
anomalies,
correctly contained in \cssnca.
On the other hand, if we evaluate by square completion
the path integral so constructed,
we obtain by consistency the form of the
gravitational action induced by the free scalar superfield
\eqn\igafsf{\eqalign{
&\e^{- I_{f}[E_M{}^A]}  = \int ({\cal D} X)_E \
\e^{-S_{f}[X, E_M{}^A]} \cr
&I_{f}[E_M{}^A] =
I[E_M{}^A; {3 / 2}, 1,a_{f}].\cr}}
In this specific computation $a_{f}= {3 \over 2}$,
but, as already mentioned,  $a_{f}$ multiplies
a local term and its value can be changed at will by adding  a local
counterterm of the same form. There is no natural choice for it, as
can be seen by rederiving $I_{f}[E_M{}^A]$
using the $c=\bar c=1$ system described by \scs\ with
$\beta_1={i\over{\sqrt 6}},\ \beta_2 = 0$. Proceeding
once more as described above
gives  $a_{f}= 1$ and shows that the value of $a_{f}$
is not uniquely determined by the symmetries of the model.
Armed with the  knowledge of $I_{f}[E_M{}^A]$, it is immediate to
prove eq. \cssea\ in full generality.
This equation is the $(1,0)$  supersymmetric
generalization of the Leutweyler action \effac.
In the superconformal gauge  defined by
\eqn\csscg{ E_+{}^M = \exp \biggl ( {i\over 2} L - {1\over 2} W \biggr ) \hat
E_+{}^M ,\ \ \  E_={}^M = \exp( -i L -  W) \hat
E_={}^M  }
it produces the following  extension of the  super-Liouville action
\eqn\cssclm{\eqalign{ S[L,W, \hat  E_M{}^A]  &\equiv
I[E_M{}^A] - I [\hat E_M{}^A]
\cr &= {1\over 24 \pi} \int d^3Z \hat e_+ \
\bigl [W \hat \nabla_+ \hat \nabla_= W ( c+ \bar c + 2a)
       + L \hat \nabla_+ \hat \nabla_= L ( 2a  -c - \bar c )  \cr & \ \ \ \
      + L \hat \nabla_+ \hat \nabla_= W   2i ( \bar c -  c )
+ 2  \hat R_1^+ W  (c+a)  + 2  \hat R_2^+ W (\bar c + a)   \cr & \ \ \ \
+ 2i \hat R_1^+ L (a-c)    + 2 i \hat R_2^+ L (\bar c - a)     \bigl ].
\cr}}
We now discuss the quantization of this $(1,0)$ chiral induced
supergravity. First of all we have to consider the
inclusion of the super-reparametrization ghosts.
It is well-known that such a ghost system has $c_{gh}=-15$
and $\bar c_{gh}= -26$, so that we need to add to the matter induced action
the following ghost contribution
\eqn\sghocont{I_{gh}[E_M{}^A] = I[E_M{}^A; -15, -26,a_{gh}].}
A second thing we have to take into account is the correct
path integral measure required for quantization.
To path integrate
over the Lorentz and Weyl fields we have to use the measures
obtained from the super-reparametrization invariant norm
on the space of worldsheet supervielbeins. The latter induces norms
for the Lorentz and Weyl fields which are
 similar to the one for  a scalar superfield
\eqn\snoruno{\eqalign{
&|| \delta L||_{E}^2 =\int d^3Z e_+ (\delta  L)^2 =
\int d^3Z \hat e_+ \exp \biggl ( {i\over 2} L +{3\over 2} W
\biggr )(\delta  L)^2
\rightarrow ({\cal D} L)_{E} \cr
&|| \delta W||_{E}^2 =\int d^3Z e_+ (\delta  W)^2 =
\int d^3Z \hat e_+ \exp \biggl ( {i\over 2} L + {3\over 2} W
\biggr )(\delta  W)^2
\rightarrow ({\cal D} W)_{E} .\cr}}
However, the path integral measures $ ({\cal D} L)_{E}$ and
$({\cal D} W)_{E}$
are not translational
invariant since the superdeterminant
$  e_+$ depends on the Lorentz and Weyl fields
themselves. It is not clear how to use these measures
to compute directly  the path integral.
Instead, one can use the translational
invariant measures obtained from
\eqn\snordue{\eqalign{
&|| \delta L||_{\hat E}^2 =\int d^3Z \hat e_+ (\delta  L)^2
\rightarrow ({\cal D} L)_{\hat E} \cr
&|| \delta W||_{\hat E}^2 =\int d^3Z \hat e_+ (\delta  W)^2
\rightarrow ({\cal D} W)_{\hat E} .\cr}}
These translational invariant measures are related to the
super-reparametrization  invariant ones by a jacobian factor
\eqn\sjac{({\cal D} L)_{ E} = J({\cal D} L)_{\hat E} ,\ \ \
({\cal D} W)_{ E} = J ({\cal D} W)_{\hat E} }
formally given by
$J = {\rm sdet} \ \exp \bigr (  {i\over 2} L +{3\over 2} W \bigl )$.
We compute $J$ in a way similar to the bosonic case treated in refs. \MM\
and briefly reviewed in the previous section.
Accordingly, we should calculate the jacobian $J$ using
the heat-kernel regularization corresponding to the \lq\lq laplacian"
$\nabla_+  \nabla_= $ which is the kinetic operator for a scalar superfield
(with the necessary modification to obtain a truly
elliptic differential operator).
It is clear that such a computation is identical
to the one that must be done to derive the gravitational action induced
by a free scalar superfield, where all of the anomalous dependence
on the Lorentz and Weyl modes of the supervielbein  is contained in the
path integral measure \duedis. In fact, one could derive these anomalies by
computing the heat-kernel regulated jacobian  corresponding to
infinitesimal symmetry transformations and obtain the induced action by
integrating the corresponding
anomalous Ward identities. The only difference
between such an induced  action and our  jacobian $J$ is
the overall normalization. Recalling eq. \igafsf, we can immediately
write down the result
\eqn\result{ J= \exp \bigl ( - (I[E_M{}^A; {3 / 2}, 1,a_{J}] -
I[\hat E_M{}^A; {3 / 2}, 1,a_{J}] ) \bigr ) }
where the last term in the exponent is a
constant included to  normalize $J$ to unity for $L=W=0$.
As could have been expected, the expression in the exponent
has the structure of the Lorentz-Weyl action in \cssclm.
Note that there is no canonical value for the parameter $a_{J}$.
To summarize, we collect in one expression all terms
containing a Lorentz and Weyl dependence
\eqn\recap{ \eqalign {I'[E_M{}^A] &\equiv  I[E_M{}^A] + I_{gh}[E_M{}^A]
-2 {\rm log} J \cr &=
I[E_M{}^A;c, \bar c,a] + I[E_M{}^A;-15,-26,a_{gh}] \cr & \ \ +
 2 (I[E_M{}^A;{3 / 2}, 1,a_{J}] - I[\hat E_M{}^A;{3 / 2}, 1,a_{J}]) \cr}}
where on the right hand side of the top line
the first term is due to the inducing matter, the second to the
ghosts for the conformal gauge
and the third one to the jacobians in \sjac.
This action can be rewritten in a more compact way as
\eqn\compact{I'[E_M{}^A] =
I[E_M{}^A;c', \bar c',a'] -2 I[\hat E_M{}^A;{3 / 2}, 1,a_{J}]}
where
\eqn\qosc{\eqalign{
& c'= c -15 +3 \cr & \bar c' = \bar c- 26 +2 \cr & a' = a + a_{gh} +
2 a_{J}. }}
It should be quantized with the translational invariant measures
\snordue. We will consider first the  quantization of the  Lorentz field,
as already explained in the bosonic case.
To extract from \compact\ the part depending on the Lorentz mode we set
\eqn\pcg{ E_+{}^M = \exp \biggl ({i\over 2} L \biggr ) \tilde E_+{}^M,\ \ \
E_={}^M = \exp(- i L ) \tilde E_={}^M,}
where the vielbein $\tilde  E_A{}^M$ now contains the Weyl field $W$,
and obtain from the relevant piece of \compact\ the following Lorentz action
\eqn\cssla{\eqalign{
&S_{Lor}[L, \tilde  E_M{}^A]  \equiv
I[E_M{}^A;c', \bar c',a'] -I[\tilde E_M{}^A;c', \bar c',a']
\cr & \  ={1\over 24 \pi} \int d^3Z \tilde e_+
\bigl [  L \tilde \nabla_+ \tilde \nabla_= L (2a'-c'- \bar c')
+ 2i \tilde R_1^+ L (a'-c') + 2i \tilde R_2^+ L(\bar c' - a') \bigl ].\cr}}
Clearly, the Lorentz field behaves as a $(1,0)$ chiral boson,
and since we have already taken care of the nontrivial part of
the path integral measure, we can immediately compute
\eqn\inte{\eqalign{&\e^{- I_{Lor}[\tilde E_M{}^A]}  = \int
({\cal D} L)_{\tilde E} \
\e^{-S_{Lor}[L,\tilde E_M{}^A]} \cr
&I_{Lor}[\tilde E_M{}^A] =
I[\tilde E_M{}^A; c_{Lor}, \bar  c_{Lor},a_{Lor}]\cr}}
with the values of the central charges given by
\eqn\ntrecc{\eqalign {
& c_{Lor}=
{3\over 2} - {{(c'-a')^2} \over {c' +\bar c' - 2a'}}
\cr & \bar c_{Lor}=
1 - {(\bar c' -a')^2 \over {c' +\bar c' - 2a'}}. \cr}}
Here above we have used for simplicity the measure $({\cal D} L)_{\tilde E}$
defined in the obvious way. Clearly the relation between this measure
and the ones previously defined  is as follows
\eqn\relmes{ \eqalign{ &({\cal D} L)_{ E} = J_1 ({\cal D} L)_{\tilde E}, \ \
\
J_1 = \exp \bigl ( - (I[E_M{}^A; {3 / 2}, 1,a_{J}] -
I[\tilde E_M{}^A; {3 / 2}, 1,a_{J}] ) \bigr ) \cr
&({\cal D} L)_{\tilde E} = J_2 ({\cal D} L)_{\hat E}, \ \ \
J_2 = \exp \bigl ( - (I[\tilde E_M{}^A; {3 / 2}, 1,a_{J}] -
I[\hat E_M{}^A; {3 / 2}, 1,a_{J}] ) \bigr ). \cr }}
Of course, $ J= J_1 J_2$. Note that the
values of the Lorentz central charges depend on the arbitrary parameter
$a'$, in a way similar to the bosonic case.
Note also that there is no particular value  of
$a_{Lor}$ which looks natural, so far.
However, we will soon discover that the Ward identities for background
Lorentz and Weyl invariance will fix a unique value for such a constant.
Now, we are left to quantize the Weyl field. We complete
the specification of the conformal gauge by setting
\eqn\fcg{ \tilde E_+{}^M = \exp \biggl (-{1\over 2}W \biggr ) \hat E_+{}^M,
\ \  \tilde E_={}^M = \exp(- W ) \hat E_={}^M}
and collect all the terms left over after the Lorentz integration
\eqn\leftover{ {I''[\tilde E_M{}^A] \equiv  I[\tilde E_M{}^A] +
I_{gh}[\tilde E_M{}^A] + I_{Lor}[\tilde E_M{}^A]
- {\rm log} J_2.
}}
There is only one jacobian here since the one for the Lorentz field
was used to construct the measure $({\cal D} L)_{\tilde E}$ in \inte\
and is effectively incorporated in $I_{Lor}[\tilde E_M{}^A]$.
We rewrite the action \leftover\ in a compact form
\eqn\compwey{I''[\tilde E_M{}^A] =
I[\tilde E_M{}^A;c'', \bar c'',a''] - I[\hat E_M{}^A;{3 / 2}, 1,a_{J}]}
where
\eqn\ccc{\eqalign{&c''= c - 15 + c_{Lor}  +{3\over 2}\cr
&\bar c''= \bar c -26 + \bar c_{Lor}  +1\cr &a'' =
a+a_{gh}+a_{Lor}+a_{J},\cr}}
and use it to obtain the following super-Liouville action
\eqn\treli{\eqalign{
&S_{Liou}[W,\hat E] \equiv I[\tilde E_M{}^A;c'', \bar c'',a''] -
I[\hat E_M{}^A;c'', \bar c'',a''] \cr
& \   ={1\over 24 \pi} \int d^3Z \hat e_+
\bigl [ W \hat \nabla_+ \hat \nabla_= W
 (c''+\bar c''+2a'' )+ 2\hat R_1^+ W (c''+ a'')
+ 2 \hat R_2^+ W (\bar c''+ a'') \bigl ]. \cr}}
It is immediate to  quantize it by evaluating the corresponding path integral
\eqn\intew{\eqalign{&\e^{- I_{Liou}[\hat E_M{}^A]}  = \int
({\cal D} L)_{\hat W} \
\e^{-S_{Liou}[W,\hat E_M{}^A]} \cr
&I_{Liou}[\hat E_M{}^A] =
I[\hat E_M{}^A; c_{Liou}, \bar  c_{Liou},a_{Liou}]\cr}}
where the values of the central charges are  given by
\eqn\ntrecc{\eqalign {
& c_{Liou}=
{3\over 2} - {{(c''+a'')^2} \over {c'' +\bar c'' + 2a''}}
\cr & \bar c_{Liou}=
1 - {(\bar c''+a'')^2 \over {c'' +\bar c'' +2a''}}. \cr}}
Plugging the values of the various central charges
 in these formulae, one can verify that
 we can achieve
\eqn\susycc{
c+ c_{Lor} -15 +c_{Liou} = \bar c+ \bar c_{Lor} -26 +\bar c_{Liou} =0}
by suitably choosing $a_{Lor}$ in $a'' $.
Let's see this in a more detailed way. One
 can first check that  $\bar c''= c''$. Then, fixing $a_{Lor}$
in $a''$ to obtain $a''= c''$ gives
\eqn\quno{\eqalign { &c_{Liou}= {3\over 2} - c'' \cr
&\bar c_{Liou}= 1 - \bar c'', \cr}}
and, consequently, eq. \susycc.
Thus, there is a unique value for $a_{Lor}$
consistent with the Lorentz and Weyl background invariance, namely
$a_{Lor}= c''-(a+a_{gh}+a_{J})$. Note that with
this specific value of $a_{Lor}$ the Liouville action
\treli\ takes the following simple form
\eqn\trelidue{ S_{Liou}  ={c''\over 6 \pi} \int d^3Z \hat e_+
\bigl [ W \hat \nabla_+ \hat \nabla_= W
+ \hat R^+ W  \bigl ]. }
 A final thing to take care of is to
fix  $a_{Liou}= -(a + a_{gh} +a_{Lor})$ in \intew\
to recover the full background Lorentz
and Weyl invariance, so that we obtain
\eqn\jatte{ I[\hat E_M{}^A]+I_{Lor}[\hat E_M{}^A]+
I_{gh}[\hat E_M{}^A]+I_{Liou}[\hat E_M{}^A] =0.}

We have thus seen that for the $(1,0)$ induced supergravity
the quantization of the Lorentz-Weyl theory gives central charges
$c_{grav}= 15 - c$ and $\bar c_{grav}=26 - \bar c $,
together with a unique value for the couplings $a_{Lor}$ and $a_{Liou}$.
These results are necessary to show the consistency of
the quantization of the supergravitational sector,
since the background Lorentz and Weyl invariance
corresponds to a change of the gauge slice.
The analysis has been very similar to the one carried out in the bosonic
case, even though the chiral nature of $(1,0)$ superspace
has made things slightly more complicated.


\newsec{Induced $\bf (1,1)$ chiral supergravity}

The $(1,1)$ superspace is the natural arena for the description of
the  spinning string  as long as one does not need
fermionic vertex operators \ref\FMS{
D. Friedan, E. Martinec and S. Shenker, Nucl. Phys. B271 (1986) 348.}.
Its non-chiral structure makes it easier to discuss the Lorentz and
Weyl anomalous action for supergravity induced by chiral supermatter
as well as its quantization, and we can follow closely
 the case of non-critical bosonic strings.

We start  with a bosonic superfield $X$
coupled to the chiral background scalars $R_1$ and $R_2$ (see appendix C)
\eqn\scb{ S[X,E_M{}^A] = {1\over { 2 \pi}} \int d^4Z E ( \nabla_+ X
\nabla_- X + \beta_1 R_1 X + \beta_2 R_2 X). }
It describes a chiral boson in superspace as can be checked by computing its
stress tensor $T_A{}^B$  defined by
\eqn\ssst{ \delta S = - {1\over \pi} \int d^4Z E \ T_A{}^B
 H_B{}^A (-1)^A,}
with $H_A{}^B$ running over the set of the six  independent variations $(
H_\pm{}^\pp, H_\pm{}^=, H_+{}^+, H_-{}^-)$.
Using the list of the dependent variations given in  appendix C,
one gets the following non-vanishing  components
of the stress tensor
evaluated on-shell and in flat superspace
\eqn\sstress{\eqalign{ T &\equiv T_\pp{}^-
= -{1\over 2} D_+ X \partial_\pp X - {1 \over 2}
\beta_1 \partial_\pp D_+ X,  \ \
c = {3\over 2} + 3 \beta_1^2, \cr
\bar T &\equiv T_\mm{}^+
= {1\over 2} D_- X \partial_\mm X
+ {1 \over 2} \beta_2 \partial_\mm D_- X,  \ \ \ \ \ \
\bar c = {3\over 2} + 3 \beta_2^2. \cr}}
We have reported also the central charges of the
corresponding super-Virasoro algebras. Their values
can be checked using the propagator
\eqn\sprop{ \langle X(Z_1) X(Z_2) \rangle = - \log (\mu^2 z_{12}^\pp
z_{12}^= ), }
where $z_{12}^\pp= z_1^\pp - z_2^\pp - \theta_1^+ \theta_2^+$
and  $z_{12}^\mm= z_1^\mm - z_2^\mm - \theta_1^- \theta_2^-$,
and computing the relevant operator product expansions.
This is enough to make sure that eq. \scb\ describes a chiral system.
We now use this chiral boson to represent superconformal systems with
arbitrary central charges $c$ and $\bar c$.
After recalling the form of the effective action due to a $c=\bar c$ system
\ref\Gr{M. Grisaru and R. Xu, Phys. Lett. B205 (1988) 486.}
\eqn\ssnca{ I_0[E_M{}^A;c]  =  {c\over 24 \pi} \int d^4Z E
  R {1\over \nabla_+\nabla_-} R,}
it is a simple task to obtain the Lorentz and Weyl anomalous
effective action induced by chiral superconformal
systems. It reads
\eqn\ssea{ I[E_M{}^A]= I[E_M{}^A;c,\bar c,a]
\equiv {1\over 24\pi}\int d^4Z E
\biggl ( c R_1 {1\over \nabla_+ \nabla_-} R_1 +
\bar c R_2 {1\over \nabla_+  \nabla_-} R_2
+ 2a \O_+ \O_- \biggr ).}
In the superconformal gauge
\eqn\sscg{ E_\pm{}^M = \exp \biggl (\pm {i\over 2} L - {1\over 2} W
\biggr ) \hat
E_\pm{}^M, }
it generates the following action
for the Lorentz and Weyl modes $L$ and $W$ of the supervielbein
\eqn\ssclm{ \eqalign{ S[L,W] &= I[E_M{}^A] - I [\hat E_M{}^A]
\cr & ={1\over 24 \pi} \int d^4Z \hat E \
\bigl [W \hat \nabla_+  \hat \nabla_- W ( c+ \bar c + 2a)
       + L \hat \nabla_+ \hat \nabla_- L ( 2a  -c - \bar c )  \cr & \ \ \ \
     +  L \hat \nabla_+ \hat \nabla_- W  2i ( \bar c -  c )
  + 2  \hat R_1 W (c+a)  + 2  \hat R_2 W (\bar c + a)   \cr & \ \ \ \
 + 2i \hat R_1 L (a-c)   + 2 i \hat R_2 L (\bar c - a)     \bigl ].
\cr}}
Again we discuss the quantization of this
$(1,1)$ supersymmetric Lorentz-Weyl action by
taking care of the Lorentz field first. To do this in a simple way,
we re-insert the Weyl field $W$ back into the background,
i.e. we set
$E_\pm{}^M = \exp(\pm {i\over 2} L ) \tilde
E_\pm{}^M$ in \ssea, and obtain
\eqn\tressla{\eqalign{
&S_{Lor}[L,\tilde E_M{}^A] \equiv I[E_M{}^A]- I[\tilde E_M{}^A]\cr
& \ = {1\over 24 \pi} \int d^4Z \tilde E \
\bigl [ L \tilde \nabla_+ \tilde \nabla_- L ( 2a  -c - \bar c )
 + 2i \tilde R_1 L (a-c)   + 2 i \tilde R_2 L (\bar c - a) \bigl ].\cr}}
Path integrating over $L$ gives
\eqn\ssinte{\eqalign{&\e^{- I_{Lor}[\tilde E_M{}^A]}  = \int
({\cal D} L)_{\tilde E} \
\e^{-S_{Lor}[L,\tilde E_M{}^A]} \cr
&I_{Lor}[\tilde E_M{}^A] =
I[\tilde E_M{}^A; c_{Lor}, \bar  c_{Lor},a_{Lor}]\cr}}
where
\eqn\ntrecc{ c_{Lor}= {3\over 2} - {{(c-a)^2} \over {c +\bar c - 2a}}, \ \ \
 \bar c_{Lor}=  {3\over 2} - {(\bar c -a)^2 \over {c +\bar c - 2a}}, \ \ \
 a_{Lor}= {3\over 2}+ {(c-a) (\bar c -a) \over {c +\bar c - 2a}}. }
Note that we have used the measure  derived from
\eqn\messs{ || \delta L||_{E}^2 =\int d^4Z E (\delta  L)^2
\rightarrow ({\cal D} L)_{E}}
which satisfies $({\cal D} L)_{E} = ({\cal D} L)_{\tilde E} $
because the superdeterminant of the vielbein in the $(1,1)$ superspace
is Lorentz invariant. As a consequence of such a quantization we obtain
\eqn\match{ c+c_{Lor} = \bar c+ \bar c_{Lor} = a+a_{Lor} = {3\over 2} +
{{c \bar c - a^2 }\over {c +\bar c - 2a}}}
where the coupling  $a$ is left arbitrary.
Now, it remains to quantize the Weyl mode
and this can be done by repeating the
analysis of \ref\kawai{J. Distler,
Z. Hlousek and H. Kawai,
Int. J. Mod. Phys. A5 (1990) 391.},
which implies that the total central charges vanish
\eqn\zero{ \eqalign{ &c_{tot}= c + c_{Lor} - 15 + c_{Liou} =0,\cr
& \bar c_{tot}=\bar c + \bar c_{Lor} - 15 + \bar c_{Liou} =0.\cr}}
We will omit the description of such an analysis here since
it is very similar to the one given in section 2 and is anyway reported
in ref. \kawai.

\newsec{Conclusions}

We have derived supergravitational actions induced by superconformal
chiral matter, namely the $(1,0)$ and $(1,1)$ chiral induced supergravities
given by eqs. \cssea\ and \ssea, respectively.
These actions are the starting points to study off-critical
heterotic and spinning strings. The chiral structure is particularly
natural in the heterotic case since the heterotic string is chiral by its own
nature (the constraint algebra which defines it,
super-Virasoro$\otimes$Virasoro, is chiral).
A  main tool in the derivation of these  chiral gravitational actions  was
the chiral boson introduced in ref. \fiore.
It consists of a  scalar (super)field
with  linear couplings to the chiral background curvature scalars,
whose effect is to induce  improvement terms in the stress tensor, so that
arbitrary central charges $ c $ and $\bar c$ can be obtained.
The induced gravitational actions we have found have an interesting chiral
structure.
In the conformal gauge they give rise to generalizations of the Liouville
action and give dynamics to both the Lorentz and Weyl modes of the vielbein.
In a stepwise quantization, the Lorentz mode, which we quantize first,
is seen to behave as a chiral boson with the effect of leveling up the
chiral mismatch between the matter central charges.
The Weyl mode is quantized afterwards,
using the free field approach of David, Distler and Kawai.
We have investigated  only local properties of the model.
Global properties are more subtle to analyze, but essential for a derivation
of critical exponents along the line of reasoning of refs. \Dav \DK.
An example of a global effect is the following one. It can be seen that the
Lorentz field acquires superselection rules, derived in the bosonic case by
Myers and Periwal using the $SL(2,C)$ invariance on the sphere
as well as factorization for more complicated surfaces \Myers.
Such rules arise because winding sector gets excited when  the
Lorentz field is  in the presence of a non-trivial topology (i.e. higher
genus
surfaces or punctures in the sphere). Similar superselection rules
can also be derived in the supersymmetric cases using the proper
supersymmetric generalization of the $SL(2,C)$ invariance group.
Another property worth of analysis is unitarity.
In fact, as noticed in \Myers, an interesting feature that arises
from the quantization of chiral non-critical string is
that the barrier $c\leq1$ of non-chiral non-critical bosonic string
can be avoided by properly choosing the coupling $a$ (see eq. \lev).
However, the Lorentz field has imaginary coupling (at least in a
euclidean signature for the worldsheet) and its central charges can become
negative for certain values of $a$. This suggest that one
should pay due attention to
unitarity in a complete theory of chiral off-critical strings.
Thus, it is clear that additional work is required to fully understand
the quantization of these induced  chiral (super)gravities.
Certainly, they describe fascinating models
which may teach us many useful lessons in string theory, chiral models
and 2d gravity.

\vfill
\eject


\appendix{A}{$\bf (0,0)$ superspace}

We describe the local geometry of the two dimensional
$(0,0)$ superspace, that is to say
of an usual Riemann surface, in such a way that immediately
generalizes to the
$(1,0)$ and $(1,1)$ superspaces.
First of all, we denote the complex coordinates of the 2d flat space
by $x^\mu= ( x^\pp, x^\mm )$ and their corresponding derivatives by
$\partial_\mu =( \partial_\pp, \partial_\mm )$.
We also use the integration
measure $d^2 x= dx^1 dx^2$, where $x^\pp = x^1 + i x^2$
and $x^\mm = x^1 - i x^2$.

A curved space is obtained by introducing
Lorentz covariant derivatives, defined as
\eqn\Clcd{ \nabla_a = e_a{}^\mu \partial_\mu + \o_a J,\ \ \ \ \ \ \ \
a=(\pp,\mm)} and  constrained by
\eqn\Ccon{ \lbrack \nabla_\pp, \nabla_= \rbrack = - R J,}
where $a$ denotes  Lorentz (flat)
indices\footnote{$^\dagger$}{The curved index $\mu$
will always appear contracted in covariant formulas and thus no confusion
should arise with the use of $(\pp,\mm)$  as flat indices.
When choosing the conformal gauge, $e_\mu{}^a = \delta_\mu{}^a$,
flat and curved indices gets identified. This explains our notation.}
and $J$ is the Lorentz
generator ( $[J,v_\pp] = v_\pp$, $[J,v_=] = -v_=$, etc.).
The Lorentz metric $\eta_{ab}$ has
$\eta_{\pp \mm} = \eta_{\mm \pp} = {1\over 2}$ as the only
 non-zero components.
The constraint in \Ccon\ is solved for the spin connection
$\o_a $ as a function of the vielbein
$e_a{}^\mu$
\eqn\Cspin{ \o_\pp = - {1\over e} \partial_\mu (e e_\pp{}^\mu ),\ \
\o_\mm =  {1\over e} \partial_\mu (e e_\mm{}^\mu ) ,}
where $e \equiv {\rm det}\  e_\mu{}^a$ and  $e_\mu{}^a$ inverse of
$e_a{}^\mu$.
The definition of the
scalar curvature $R$ then gives
($e_a \equiv e_a{}^\mu \partial_\mu$)
\eqn\CR{\eqalign {&R= R_1 +R_2,\cr
&R_1 =-\nabla_\pp \o_= = -(e_\pp -\o_\pp) \o_=,\cr
&R_2 =\nabla_= \o_\pp = (e_= +\o_=) \o_\pp .\cr }}
The scaling properties
under Weyl ($\s$) and Lorentz ($\l$) transformations, given by
\eqn\csca{
  e_\mu{}^{\pp} \rightarrow \exp({\s - i\l} )  e_\mu{}^{\pp} ,\ \ \ \
e_\mu{}^{\mm} \rightarrow \exp({\s + i\l})   e_\mu{}^{\mm} }
are easily derived from the   above formulas, and read
\eqn\Ctr{ \eqalign{ &\o_{\pp} \rightarrow  \exp({-\s + i \l})
\bigl ( \o_\pp - \nabla_\pp ( \s + i\l) \bigr ), \ \ \ \
\o_{\mm} \rightarrow  \exp({-\s - i \l} )
\bigl ( \o_\mm + \nabla_\mm ( \s - i\l) \bigr ), \cr
& e \rightarrow \exp(2\sigma) e, \ \ \ \
  e  R_1 \rightarrow
e \bigr ( R_1 - \sq (\sigma -i\lambda )\bigl ), \ \ \ \
e R_2 \rightarrow  e \bigr
( R_2 - \sq (\sigma + i\lambda )\bigl ),  \cr
&e  R \rightarrow
e \bigr ( R - 2\sq \sigma \bigl ) , \cr
}}
where $\sq \equiv \nabla_= \nabla_\pp = \nabla_\pp \nabla_=$
 is the laplacian acting on scalars. Note that $R$ is Lorentz invariant
and coincides, up to some normalization, with  standard definitions
which make use of the metric tensor
$g_{\mu\nu} = e_\mu{}^a e_\nu{}^b \eta_{ab} $. With our
normalization the Euler theorem for a surface of genus $g$
reads: $ \int d^2x e  R = (2- 2g) \pi$.


\appendix{B}{$\bf (1,0)$ superspace}

The rigid $(1,0)$ superspace is described by the coordinates
$Z^M =( \theta^+, x^\pp, x^\mm )$, with $\theta^+$  fermionic
and $ x^\pp, x^\mm $ complex conjugate bosonic  coordinates.
The  covariant derivatives of rigid superspace are denoted collectively by
$D_M=( D_+, \partial_\pp, \partial_\mm ) $,
with
\eqn\Asder{ D_+ = {\partial \over {\partial \theta^+}} +  \theta^+
{\partial \over{\partial x^\pp }}}
and the only non-trivial graded commutator is
$\{ D_+,D_+\} = 2 \partial_\pp$. The integration measure is denoted by
$d^3Z= d^2x d\theta^+$.

The geometry of $(1,0)$  supergravity is locally
defined by introducing Lorentz covariant derivatives
\eqn\Ader{ \nabla_A = E_A{}^M D_M + \O_A J, \ \ \ \ \ \ \ \ A=(+,\pp,\mm)}
constrained by
\eqn\Acon{ \eqalign{
 \{ \nabla_+, \nabla_+ \} &= 2 \nabla_{\pp} \cr
 [ \nabla_+, \nabla_{\mm} ] &= R^+ J. \cr}}
Of the $9+3$  superfields  contained in the supervielbein and spin
connection, $7$ are killed by the constraints, leaving
$5$ independent superfields,
usually called prepotentials, describing $(1,0)$ supergravity.
We will not need the explicit solution
of the constraints  in terms of the prepotentials, which are described in
\GG.
It will be  enough to note that the first constraint in \Acon\
gives $E_\pp{}^M$ and $\O_\pp$ as functions of $E_+{}^M$ and $\O_+$,
while the second one allows
in particular to solve for the remaining components
of the spin connections
\eqn\Ascon{ \O_+ = - {1\over e_+} D_M (e_+ E_+{}^M), \ \ \
\O_\mm = (-1)^{M} {2\over e_+} D_M (e_+ E_\mm{}^M) ,}
where $ e_+ = {\rm sdet} E_M{}^A$, with
$ E_M{}^A$ inverse of $E_A{}^M$.
Note that the superdeterminant $e_+$ transforms under Lorentz rotations as
indicated by the Lorentz index. It is a bosonic object.
 The second equation in \Acon\ gives the definition
of the curvature $R^+$, which can be naturally split in chiral pieces
\eqn\AsR{ \eqalign{ &R^+= R^+_1 +R^+_2,\cr
&R^+_1 \equiv \nabla_+ \O_\mm = \bigl (E_+ -\O_+\bigr ) \O_\mm ,\cr
&R^+_2 \equiv - \nabla_\mm \O_+ = -\biggl (E_\mm +{1\over2}\O_\mm\biggr )
\O_+ ,\cr}}
where $E_A \equiv E_A{}^M D_M$.
The constraints describing the local $(1,0)$ superspace are
manifestly super-reparametrization and local Lorentz covariant,
i.e. covariant under
\eqn\Asym{\nabla_A \rightarrow \hat \nabla_A =
\e^{K + \Lambda} \nabla_A \e^{-K-\Lambda}, }
with $K \equiv K^M D_M $ and $\Lambda \equiv iLJ $.
The superfields $ K^M$ and $L$
describe  super-reparametrizations and Lorentz rotations, respectively.
In addition,  Weyl transformations can be defined   as
\ref\Howe{P. Howe and R. Tucker, Phys. Lett. B80 (1978) 138.}
\eqn\weyl{ \eqalign{
E_+{}^M & \rightarrow \hat E_+{}^M{} = \exp \biggl ({ - {1\over 2}W }
\biggr )  E_+{}^M ,\cr
E_\mm{}^M &\rightarrow \hat E_\mm{}^M{} = \exp({ - W })  E_\mm{}^M. \cr}}
The Weyl transformation
rules on the other components of the supervielbein and
on the spin connection
follow  form the constraints.
Altogether there are  5  local symmetries which are enough to
locally gauge fix
the full supergeometry to  a given background value.
 The Weyl and Lorentz transformation on the relevant
geometrical objects needed in the text are as follows
\eqn\Awl{\eqalign{
E_+{}^M & \rightarrow  \exp \biggl ({{i\over 2} L - {1\over 2}W }
\biggr )  E_+{}^M , \ \ \
E_\mm{}^M \rightarrow \exp ({-i L - W } ) E_\mm{}^M ,\cr
e_+ &\rightarrow \exp  \biggl ({{i\over 2} L + {3\over 2}W } \biggr ) e_+
,\cr
\O_+ &\rightarrow \exp \biggl ({{i\over 2} L - {1\over 2}W } \biggr )
\bigl ( \O_+ - E_+ (iL+W) \bigr) ,\cr
\O_= &\rightarrow \exp(-iL -W )\bigl ( \O_= - E_= (iL-W) \bigr) ,\cr
e_+ R_1^+ &\rightarrow e_+  \bigl ( R_1^+ -\nabla_+ \nabla_= ( iL - W)
\bigr),
\ \ \
e_+ R_2^+ \rightarrow e_+ \bigl ( R_2^+  +\nabla_+ \nabla_= ( iL + W)
\bigr) , \cr
e_+ R^+ &\rightarrow  e_+  \bigl ( R^+  + 2 \nabla_+  \nabla_=
 W \bigr) .\cr }}
To derive stress tensors  form superspace actions,
we need  to vary the  vielbein. Of course, not all variations are
independent because of the constraints.
Denoting the vielbein variations
by $H_A{}^B = \delta E_A{}^M E_M{}^B $
and varying the torsion constraints which follow from \Acon, one notices
that  the 5 variations $ (H_+{}^A, H_={}^=, H_={}^\pp)$
can be taken as independent, while  other  variations are as follows
\eqn\Adep{\eqalign{
H_\pp{}^+ &= \nabla_+ H_+{}^+ +{1\over 2} (\nabla_+ H_={}^= -
             \nabla_= H_+{}^=) \cr
H_\pp{}^\pp &= 2H_+{}^+ + \nabla_+ H_+{}^\pp \cr
H_\pp{}^\mm &= \nabla_+ H_+{}^= \cr
H_\mm{}^+ &= {1\over 2} ( \nabla_+ H_={}^\pp -
             \nabla_= H_+{}^\pp )\cr
\delta \O_+ &= \nabla_+ H_={}^= -\nabla_= H_+{}^= + H_+{}^A\O_A \cr
\delta \O_\mm &= \nabla_\pp H_={}^\pp -2 \nabla_= H_+{}^+ - \nabla_+ \nabla_=
H_+{}^\pp  + H_={}^A\O_A +H_+{}^\pp R^+ .\cr}}
It is immediate to derive also
\eqn\Adv{ \delta e_+ = - e_+ ( H_+{}^+ + H_={}^= + \nabla_+ H_+{}^\pp ).}


\appendix{C}{$\bf (1,1)$ superspace}

The rigid $(1,1)$ superspace is described by the coordinates
$Z^M= ( \theta^+, \theta^-, x^\pp, x^\mm )$ and  susy covariant
derivatives
$D_M=( D_+,D_-, \partial_\pp, \partial_\mm ) $,
with
\eqn\Bcs{ D_+ = {\partial \over {\partial \theta^+}} +  \theta^+
{\partial \over{\partial x^\pp}},\ \
D_- = {\partial \over {\partial \theta^-}} +  \theta^-
{\partial \over{\partial x^=}}.}
The non-trivial graded commutators are
$\{ D_+,D_+\} = 2 \partial_\pp$ and $\{ D_-,D_-\} = 2 \partial_\mm$
while the integration measure is given by
 $d^4Z= d^2 x d\theta^- d\theta^+$.

The geometry of $(1,1)$  supergravity is defined locally
by the  Lorentz covariant derivatives
\eqn\Bder{ \nabla_A = E_A{}^M D_M + \O_A J, \ \ \ \ \ \ \ \ A=(+,-,\pp,\mm)}
constrained by
\eqn\Bcon{ \eqalign{
 \{ \nabla_+, \nabla_+ \} &= 2 \nabla_{\pp} \cr
\{ \nabla_-, \nabla_- \} &= 2 \nabla_{\mm} \cr
 \{ \nabla_+, \nabla_- \} &= R J. \cr}}
Of the $16+4$  superfields  contained in the supervielbein and spin
connection, $14$ are killed by the constraints, leaving
$6$ independent  prepotentials describing $(1,1)$ supergravity.
Again, we do not need the full solution in terms of the prepotential,
which can be found in
\ref\Martin{ M. Rocek, P. van Nieuwenhuizen and S.C. Zhang,
 Ann. Phys. (N.Y.) 172 (1986) 348. }.
It is enough to
note that the first two constraints in \Bcon\
gives $E_A{}^M$ and $\O_A$
with lower bosonic indices
as functions of those with lower fermionic indices,
while the third constraint allows to solve for the remaining component
of the spin connections as  function of the vielbein components
\eqn\Bscon{ \O_\pm = \mp{2\over E} D_M (E E_\pm{}^M), }
where $ E = {\rm sdet} E_M{}^A$ and $ E_M{}^A$ inverse of $E_A{}^M$.
The scalar  curvature $R$ can be naturally split in chiral pieces
\eqn\BsR{ \eqalign{ &R = R_1 +R_2\cr
&R_1 \equiv \nabla_+ \O_- = \biggl (E_+ - {1\over 2}\O_+\biggr ) \O_-\cr
&R_2 \equiv  \nabla_- \O_+ = \biggl (E_- +{1\over2}\O_-\biggr ) \O_+ ,\cr}}
with $E_A \equiv E_A{}^M D_M$.
The above constraints describing the local $(1,1)$ superspace are
manifestly  covariant under
\eqn\Bsym{\nabla_A \rightarrow \nabla_A'=
\e^{K +\Lambda} \nabla_A \e^{-K- \Lambda}, }
with $K\equiv K^M D_M $ and $\Lambda \equiv i L J$
describing respectively super-reparametrizations and Lorentz rotations.
In addition, Weyl transformations can be defined   by \Howe
\eqn\Bweyl{E_\pm{}^M \rightarrow
\hat E_\pm{}^M{}= \e^{-{1\over 2}W} E_\pm{}^M.}
The rules on the other components of the supervielbein and
on the spin connection
follow form the constraints.
Altogether there are  6  local symmetries which are enough to locally gauge
fix    the full supergeometry to  a given background value.
We list  the following transformation properties of some geometrical objects
under a Lorentz and Weyl transformation
\eqn\Bwl{\eqalign{
E_\pm{}^M & \rightarrow  \exp \biggl ({\pm{i\over 2} L - {1\over 2}W }
\biggr )
 E_\pm{}^M \cr
E &\rightarrow \exp(W)  E \cr
\O_\pm &\rightarrow \exp \biggl ({\pm {i\over 2}L-{1\over 2}W }
\biggr )
\bigl(\O_\pm-E_\pm(iL \pm
W) \bigr)\cr
E R_1 &\rightarrow E \bigl ( R_1
                  - \nabla_+ \nabla_- ( iL - W) \bigr) \cr
ER_2 &\rightarrow  E \bigl ( R_2
                   +\nabla_+ \nabla_- ( iL + W) \bigr) \cr
ER &\rightarrow  E \bigl ( R
                   +2 \nabla_+ \nabla_-  W \bigr) .\cr }}
To conclude, we report  the dependent variations of the vielbein, spin
connection and superdeterminant of the vielbein
as function of the $6$ independent vielbein variations
 $ (H_+{}^+, H_-{}^- , H_\pm{}^\pp, H_\pm{}^\mm )$:
\eqn\Adep{\eqalign{
H_+{}^- &= - {1\over2 } (\nabla_+ H_-{}^\mm + \nabla_- H_+{}^= ) \cr
H_\pp{}^\mm &= \nabla_+ H_+{}^\mm \cr
H_\pp{}^\pp &= 2 H_+{}^+ +  \nabla_+ H_+{}^\pp \cr
H_\pp{}^- &= -{1\over 2} (  \nabla_\pp H_-{}^=
+ \nabla_+ \nabla_- H_+{}^= - H_+{}^= R) \cr
H_\pp{}^+ &= \nabla_+ H_+{}^+ + \nabla_+ H_-{}^- -{1\over 2} (
\nabla_\mm H_+{}^=
+ \nabla_- \nabla_+ H_-{}^= - H_-{}^= R) \cr
H_-{}^+ &= - {1\over2 } (\nabla_- H_+{}^\pp + \nabla_+ H_-{}^\pp ) \cr
H_\mm{}^\pp &= \nabla_- H_-{}^\pp \cr
H_\mm{}^\mm &= 2 H_-{}^- +  \nabla_- H_-{}^\mm \cr
H_\mm{}^+ &= -{1\over 2} (  \nabla_\mm H_+{}^\pp
+ \nabla_- \nabla_+ H_-{}^\pp + H_-{}^\pp R) \cr
H_\mm{}^- &= \nabla_- H_-{}^- + \nabla_- H_+{}^+ -{1\over 2} (
\nabla_\pp H_-{}^\pp
+ \nabla_+ \nabla_- H_+{}^\pp + H_+{}^\pp R) \cr
\delta \O_+ &= 2 \nabla_+ H_-{}^- -\nabla_= H_+{}^=
- \nabla_- \nabla_+ H_-{}^=  + H_-{}^= R  + H_+{}^A\O_A \cr
\delta \O_- &= - 2 \nabla_- H_+{}^+ + \nabla_\pp H_-{}^\pp
+ \nabla_+ \nabla_- H_+{}^\pp  + H_+{}^\pp R  + H_-{}^A\O_A  \cr
 \delta E &= - E ( H_+{}^+ + H_-{}^- + \nabla_+ H_+{}^\pp
 + \nabla_- H_-{}^\mm). \cr}}

\footatend\vfill\supereject\immediate\closeout\rfile\writestoppt
\baselineskip=14pt\centerline{{\bf References}}\bigskip{\frenchspacing%
\parindent=20pt\escapechar=` \input refs.tmp\vfill\eject}\nonfrenchspacing
\end

We can perhaps note at this point that, if needed, one can
straightforwardly write down the stress tensor for the Lorentz
field. After rescaling
$ L \rightarrow L \bigl ( {12 \over { c' +\bar c' -2a'}} \bigr)^{ 1\over 2}$
to achieve a standard normalization for a chiral boson,
one  can adapt the formulas in \csstress\ to the present case
\eqn\cssL{\eqalign{ T_{Lor} &=
 -{1\over 2} D_+ L \partial_\pp L - {1 \over 2} \beta_1^{Lor}
 \partial_\pp D_+ L, \ \ \ \ \  \beta_1^{Lor} = { i (a'- c')
\over \sqrt{ 3( c' + \bar c' - 2a' )}}, \cr
\bar T_{Lor} &= -{1\over 2} \partial_\mm L  \partial_\mm L
- {1 \over 2}  \beta_2^{Lor} \partial_\mm^2  L, \ \ \ \ \ \ \ \ \ \ \
\beta_2^{Lor} ={ i ( \bar c'- a')
\over \sqrt{ 3( c' + \bar c' - 2a' )}}. \cr}}

 One could fix it by requiring
$ c +c_{Lor} = 15 $, which implies that the Weyl invariance is recovered once

the reparametrization ghosts are introduced (so that no quantization
of the Weyl mode is needed), but  we will avoid this fine tuning
and leave related comments to the next section.

Finally, we wish to clarify  our comments on the possibility of
quantizing the  Lorentz field alone, as remarked after eq. \comb\
in the bosonic case, and after eq. \match\ in the $(1,1)$ supersymmetric
case. First of all, this situation is present also for the
$(1,0)$ supergravity, though we have not mentioned it in the
corresponding section  for the sake of clarity.
We describe it now.  If the Weyl field is not
 going to be quantized, one has to drop one of the two jacobians
introduced in eq. \recap. Then, after integrating out the Lorentz
field one must set
$c+c_{gh}+c_{Lor}= \bar c+\bar c_{gh}+\bar c_{Lor}=a+a_{gh}+a_{Lor}=0$
by suitably choosing the coefficient $a$.
One may check that this is possible, thereby obtaining
a consistent string model
without the need of quantizing the Weyl mode,
as in  the critical case \ref\het{D. Gross, J. Harvey, E. Martinec and R.
Rhom,
 Nucl. Phys. B256 (1985)  253; Nucl. Phys. B267 (1986) 75.}. However,
in our stepwise quantization we have path-integrated the Lorentz field first.
In such a procedure, we completed squares and shifted the field. But recall
that the background contains the Weyl mode. This implies that shifting the
field really corresponds to taking a
linear combination of the Lorentz and Weyl modes.
When the Weyl mode is not quantized, as in the cases described just above,
one is then quantizing only this linear combination.
The need to fix the coupling $a$ can then be understood as the requirement
that the chiral induced action  be invariant under the orthogonal
linear combination of the Lorentz and Weyl modes.
In this case the geometrical picture is  somewhat obscure,
and it is perhaps more interesting to focus on the models where both
the Lorentz and Weyl mode are quantized. There is more freedom in this latter

case since  the off critical string depends parametrically on the
coupling $a$.